\documentclass[a4paper,10pt,twoside]{cpc-hepnp}

\usepackage{multicol}
\usepackage{graphicx}
\usepackage{booktabs}
\usepackage{amssymb,bm,mathrsfs,bbm,amscd}
\usepackage[tbtags]{amsmath}
\usepackage{lastpage}

\begin{document}

\fancyhead[co]{\footnotesize G. Montagna~et al: Accuracy of the Monte Carlo generators for luminosity measurements}
\footnotetext[0]{Received 9$^{th}$ February 2010}

\title{Status and accuracy of the Monte Carlo generators \\ for luminosity measurements}

\author{%
      G. Montagna$^{1,2}$
\quad G. Balossini$^{1,2}$
\quad C. Bignamini$^{1,2}$\\
\quad C. M. Carloni Calame$^{3}$
\quad O. Nicrosini$^{2}$
\quad F. Piccinini$^{2}$
}
\maketitle

\vskip 8pt
\address{%
1~Dipartimento di Fisica Nucleare e Teorica, Universit\`a di Pavia, Pavia, 27100, Italy\\
2~Istituto Nazionale di Fisica Nucleare, Sezione di Pavia, Pavia, 27100, Italy\\
\vskip -2pt
3~School of Physics \& Astronomy, University of Southampton, Southampton SO17 1BJ, U.K.
}

\begin{abstract}
The status and accuracy of the precision Monte Carlo generators used for luminosity measurements at flavour factories is reviewed. It is shown that, thanks to a 
considerable, long-term effort in tuned comparisons between the predictions of independent programs, as 
well as in the validation of the generators against the presently available calculations of the next-to-next-to-leading order 
QED corrections to Bhabha scattering, the theoretical accuracy reached by the most precise tools is of about one per mille. This error estimate is valid for realistic 
experimental cuts, appears to be quite robust and is already sufficient for very accurate luminosity measurements. However, recent progress and possible 
advances to further improve it are also discussed.
\end{abstract}

\begin{keyword}
Luminosity, QED, Bhabha scattering, radiative corrections, Monte Carlo, NNLO calculations.
\end{keyword}

\begin{pacs}
12.15.Ji, 12.20.-m, 13.40.Ks
\end{pacs}

\begin{multicols}{2}

\section{Introduction}

Precision tests of the Standard Model, indirect bounds on the Higgs boson mass and constraints on candidate models of New Physics rely on the predictive power of quantum corrections to accurately measured observables, such as the electroweak parameters and the anomalous magnetic moment of the muon 
\cite{Passera:2010ev}. However, this predictability is somewhat obscured by the non-perturbative 
strong interaction effects affecting the calculation of the light quark loop contribution to the muon anomaly and the running of the electromagnetic 
coupling constant \cite{Teubner:2010ah}. As it is well known, the most satisfactory solution to this problem consists 
in rephrasing, via dispersion relations, these non-perturbative 
corrections in terms of the experimental cross section of electron-positron ($e^+ e^-$) annihilation into hadrons measured in the region of low-lying hadronic resonances. 
This measurement, with an accuracy of the order of one per cent or even better, is an important goal of the physics program of 
the high-luminosity $e^+ e^-$ colliders of intermediately high energy, i.e. the flavour factories VEPP-2M/VEPP-2000, DA$\Phi$NE,
BEPC, CESR, KEKB and PEP-II~\cite{Actis:2009gg}. On the other hand, the measurement of the hadron production cross section, either through direct scan or 
the method of radiative return, requires, quite generally, a detailed knowledge of the 
collider luminosity~\footnote[1]{For completeness, it is worth mentioning that 
a clever alternative to the standard luminosity-based measurement of the hadronic cross section is adopted by the BaBar collaboration \cite{Davier:2010rg} and 
is under way by the KLOE experiment \cite{Muller:2009pj,Actis:2009gg}. It consists in normalizing the hadron production cross section measured through radiative return 
with radiative muon events in each bin, thus avoiding or, better, 
minimizing theoretical input and allowing for the cancellation of many common systematic uncertainties, at a price, however, 
of an increased statistical error.}. At $e^+ e^-$ accelerators, the most precise 
and widely used strategy to determine the luminosity is by means of the relation $\int\!\mathcal{L}\,{\rm d} t = N / {\epsilon \, \sigma}$, where $N$ is the number of events of 
some chosen reference process, $\epsilon$ the experimental selection efficiency and $\sigma$ the theoretical cross section of the 
normalization process. 
Therefore, the luminosity monitoring processes must be characterized by high statistics, clean topology and small amount of background, and should be calculable 
with high theoretical accuracy. This is the reason why the QED processes of Bhabha scattering, two photon and muon pair production are the normalization processes used at flavour factories \cite{Ambrosino:2006te,cleo:2007zt,asner:2008}, Bhabha scattering being of primary importance because of its large cross section. 
At all the flavour factories, the final-state particles of the above processes are selected at large scattering angles, using the same detectors employed for the measurement of the hadronic cross section. Typical experimental 
luminosity errors lie in the range between few per mille and one per cent, e.g. 0.3\%, 0.7\% and $\sim 1$\% for present Bhabha measurements at KLOE, 
BaBar and CLEO, respectively, and at the level of $\sim 0.2 \div 0.3$ for the forthcoming CMD-3 
measurement at the VEPP-2000 \cite{Fetodovich:2009}.
These experimental requirements necessitate theoretical calculations of Bhabha scattering and, more generally, of 
normalization processes with a precision 
of about one per mille, not to spoil the total accuracy in the luminosity measurement given by the 
sum in quadrature of the fractional experimental and theoretical 
uncertainty. Thanks to the effort of various groups, this high theoretical 
precision has been now successfully achieved, as discussed in the following. 

The main focus of the present contribution is on large-angle Bhabha scattering, which is the process of primary and wider interest and for which the largest part of the 
most up-to-date results is available. A complete inventory of the generators not addressed here and used at flavour factories for the simulation of QED and other 
processes can be found in \cite{Actis:2009gg}.

\section{Status and theory of the most precise luminosity generators}

During the last few years a remarkable progress occurred in reducing the theoretical error 
contribution to the luminosity measurements through the realization of newly conceived 
Monte Carlo (MC) programs or the validation of codes used in the 90s by LEP/SLC experiments 
from the $Z$ peak region down to the energy range 
of interest for flavour factories. In parallel, an impressive effort was put in the calculation of 
the next-to-next-to-leading order (NNLO) QED corrections to Bhabha scattering 
(see \cite{Actis:2009gg} for a recent review), that turned out to be (and will presumably continue to be) essential 
to assess the accuracy of the programs.

Two dedicated event generators, BabaYaga@NLO \cite{CarloniCalame:2000pz,CarloniCalame:2001pl,CarloniCalame:2003yt,Balossini:2006wc,Balossini:2008xr} and 
MCGPJ \cite{Arbuzov:2005pt,Arbuzov:1997pj,Fetodovich:2009}, were 
released in 2005-2006 to provide predictions for the cross section of the large-angle Bhabha process, as well as 
of two photon and muon pair production, with a theoretical accuracy at the level of  0.1\%. In addition, codes well-known since 
the time of LEP/SLC operation, like BHWIDE \cite{Jadach:1995nk}, were extensively used by the experimentalists in
data analysis at flavour factories. All these programs include, albeit according to different formulations, 
exact next-to-leading order (NLO) QED corrections supplemented with leading logarithmic contributions
related to multiple soft and collinear photon emission. Such ingredients, together with the vacuum polarization 
correction, are strictly necessary to achieve a theoretical 
precision down to the per mille level, as extensively proved in the literature 
\cite{Fetodovich:2009,Balossini:2006wc,Balossini:2008xr,Arbuzov:2005pt}. Indeed, when considering Bhabha scattering with typical selection cuts 
and at centre of mass (c.m.) energies from $\phi$ to $B$ factories, 
the NLO photonic corrections amount to about 15$\div$20\% (with non-logarithmically enhanced
contributions at the few per mille level), vacuum polarization introduces 
a correction of several per cent and higher-order leading effects lie between 1$\div$2\%. The common theoretical feature of 
BabaYaga@NLO, BHWIDE and MCGPJ is the appropriate combination (matching) of traditional techniques for the calculation of NLO corrections (Feynman diagrams) with universal methods for the treatment of higher-order 
contributions (QED collinear Structure Functions in MCGPJ, QED Parton Shower in BabaYaga@NLO and YFS exponentiation in BHWIDE). Consequently, the main source of theoretical uncertainty of these precision generators is due to the incomplete or approximate inclusion of NNLO QED corrections, whose
calculation comes therefore to play the role of benchmark for the accuracy of the programs. 

\section{Accuracy of the luminosity generators vs. NNLO QED calculations}

It is customary to define the total accuracy of the generators in terms of a so-called technical and 
theoretical (or physical) precision. As independent programs implement, as discussed above, radiative corrections of the 
same physical origin but according to different theoretical and computational details, this may give rise to possible differences between their predictions. 
This defines what is known as technical precision. However, the most important source of uncertainty is that 
induced by missing or partially accounted 
contributions, like e.g. NNLO terms present in the exact perturbative calculations. These missing or approximately included 
ingredients define the so-called theoretical accuracy.
During the Working Group (WG) on ``Radiative corrections and MC generators for low energies'' \cite{Actis:2009gg} 
a significant work was done to deeply investigate the size of both uncertainties. 
The results summarized in the following are the main conclusions of this effort.

\subsection{Technical precision}

The typical strategy to settle the technical precision of theoretical tools was initiated at CERN in the 90s during the 
WGs on precision physics at LEP and consists in performing tuned comparisons between the predictions of 
the codes independently developed by different groups. The word tuned means that the comparisons, to be meaningful, must be performed using the same set of parameters and cuts, in order to really test the reliability of the programs in all the technical details underlying their rather complex structure. Several tests of this type for the Bhabha process, 
according to different selection cuts and c.m. energies, are documented in \cite{Actis:2009gg}. It emerges that BabaYaga@NLO, BHWIDE and MCGPJ typically agree 
within $\sim$ 0.1\% for integrated cross sections and some per mille for distributions. Therefore, the work on tuned comparisons 
showed that the technical precision of the MC luminosity generators is well under control, the remaining differences being due to understood 
different details in the implementation of theoretical ingredients featuring the same physics (factorized vs. additive formulations, different realization of exponentiation effects). 

\subsection{Theoretical accuracy}

Concerning the theoretical accuracy of the generators, there are essentially two possible procedures to assess it. 
The first one is to compare the predictions of exact NNLO QED calculations with the corresponding $O(\alpha^2)$ 
results of the generators, if the above perturbative calculations are available 
in the literature, as it is presently true for Bhabha scattering. The second one is to estimate the size of partially accounted 
higher-order corrections, if exact or complete calculations are not yet available, or, as it would better 
to say now, were not available till recently. Actually, as further discussed in the following, the remaining ingredients 
necessary for complete NNLO predictions, unavailable until a few months ago, are now at hand: {\it i)} the full calculation of lepton and 
pion pair corrections, accounting for real pair emission beyond the soft approximation \cite{Czyz:2009}; {\it ii)} the calculation of the complete set of one-loop corrections 
to the radiative process $e^+ e^- \to \gamma^\star (s), \gamma^\star (t) \to e^+ e^-\gamma$~\cite{Actis:2009uq}. However, these ultimate ingredients 
appeared in the literature just a few weeks before the beginning of the conference, a too limited time
to perform all the phenomenological investigations necessary to obtain and present an update of the error estimate based also 
on the inclusion of these new calculations.

Considering first those NNLO corrections already under control thanks to independent calculations and 
rather deeply investigated in connection with the precision of the generators, they can be divided in three gauge-invariant classes:
\begin{enumerate}
\item photonic corrections, computed in \cite{Penin:2005eh} in the soft-photon approximation;
\item Electron loop corrections, calculated independently by two groups in 
\cite{Bonciani:2004gi,Actis:2007gi} and found to be in perfect agreement;
\item Heavy fermion and hadron loop corrections due to the insertion of $\mu$, $\tau$ leptons and hadrons in the photon propagators, 
computed in \cite{Becher:2007cu,Bonciani:2007eh,Actis:2007,Kuhn:2008zs}.
\end{enumerate}
As shown by these calculations and summarized in~\cite{Actis:2009gg}, the typical size of the above NNLO corrections to the Bhabha differential cross section is of the order of some per mille as a function of the electron scattering angle, both at $\phi$ and $B$ factories. Not surprisingly, they are largely dominated by photonic corrections and, to a less extent, by electron loop contributions. However, the bulk of such corrections is included in the generators as a bonus of the matching procedure, that allows to  effectively incorporate the most important piece of the NNLO photonic corrections, 
given by the infrared-enhanced $O(\alpha^2 L)$ terms \cite{Montagna:1996gw} ($L$ being the collinear logarihtm). Moreover, also the dominant part 
of the electron, heavy fermion and hadron NNLO corrections, originating from the reducible two-loop diagrams \cite{Kuhn:2008zs}, 
is included at $O(\alpha^2)$, albeit partially, through the inclusion of the vacuum polarization effects in the NLO 
contribution of the matched formulation of the generators. Hence, the relevant question is to what extent the exact NNLO 
results compare with the predictions of the codes at the same perturbative order.

To answer this question, detailed comparisons were carried out between the calculations \cite{Penin:2005eh,Bonciani:2004gi} and the corresponding results of BabaYaga@NLO. It turns out that, as a consequence of the factorization property of the infrared and collinear singularities, as well as of the dominance of two-loop reducible graphs within the full set of NNLO contributions, the relative deviation between the BabaYaga@NLO predictions and the results of \cite{Penin:2005eh,Bonciani:2004gi} does not exceed 2$\times 10^{-4}$, an order of magnitude below the target one per mille precision. An even smaller, completely negligible discrepancy, at the level of 1$\times 10^{-5}$, was registered between the BabaYaga@NLO calculation of the two hard photon emission process $e^+ e^- \to e^+ e^-\gamma\gamma$ (a further ingredient of the complete NNLO Bhabha cross section) and the exact diagrammatic answer~\cite{Balossini:2006wc}. 

Concerning those NNLO corrections only preliminarly scrutinized and requiring further work with relation to the accuracy of the generators, 
they are, as already remarked: 
\begin{enumerate}
\item the exact NLO soft plus virtual QED corrections to hard photon emission in full ($s$- plus $t$-channel) Bhabha scattering;
\item the complete lepton and hadron pair corrections, containing, in addition to the exact virtual and real 
soft-photon contributions, also the exact four-particle matrix elements, like 
$e^+ e^- \to e^+ e^- (l^+ l^-), l= e, \mu, \tau$, and $e^+ e^- \to e^+ e^- (\pi^+ \pi^-)$. 
The parentheses mean that those two final-state particles must be undetected to contribute to the Bhabha signature.
\end{enumerate}

The exact calculation of the one-loop corrections to $e^+ e^- \to e^+ e^-\gamma$ 
was performed recently in \cite{Actis:2009uq}, also thanks to innovative advances in the 
reduction of complex loop structures. A corresponding numerical code to derive results of 
phenomenological interest (for $\mu^+\mu^-\gamma$ production too) 
 is available as well. However, in the absence as of today of explicit comparisons between the approximate results of the generators (again, possible as a by-product of the matching) and the exact calculation, the present estimate of the theoretical uncertainty in the calculation of 
 this NNLO ingredient must necessarily rely on partial results known in the literature \cite{Jadach:1995hy,Glosser:2003ux,Jadach:2006fx}. 
 The latter analyses apply to $t$- and $s$-channel processes separately and neglect contributions, like pentagon-like diagrams, 
 taken into account in the exact one-loop calculation~\cite{Actis:2009uq}. 
 However, from the studies \cite{Jadach:1995hy,Glosser:2003ux,Jadach:2006fx} one can argue that the error induced by the NLO corrections to the radiative Bhabha process 
 conservatively amounts to about $5 \times 10^{-4}$. The uncertainty coming from the contribution of lepton and hadron pairs completes the picture of the sources of theoretical error. In this context, a DESY Zeuthen and Katowice theory collaboration \cite{Czyz:2009} recently performed a complete calculation of such contribution using all the necessary virtual and real pair ingredients discussed above. The approximate results of BabaYaga@NLO, that include part of the virtual pair corrections
  thanks to the insertion of the vacuum polarization in the NLO diagrams but does not take into account real pair effects and the two-loop pair contributions to the electron form factor, 
 were preliminary compared in \cite{Actis:2009gg} (for lepton pairs only) with the exact DESY Zeuthen-Katowice calculation. It comes out that the approximation of BabaYaga@NLO agrees rather well with the complete prediction, showing relative deviations not exceeding few units in $10^{-4}$ in the presence of realistic luminosity cuts 
 for KLOE and BaBar experiments. 

A final, parametric contribution to the theory uncertainty comes from the non-perturbative light quark contribution 
$\Delta\alpha^{(5)}_{\rm hadr}(q^2)$ to the vacuum polarization. It can be estimated by using the routines, like 
HADR5N \cite{Jegerlehner:2006ju}
and HMNT \cite{Teubner:2010ah,Hagiwara:2003da}, that parametrize this correction in terms of the low-energy hadronic cross section. Indeed these parameterizations 
 return, in addition to $\Delta\alpha^{(5)}_{\rm hadr}(q^2)$, the experimentally-driven error $\delta_{\rm hadr}$ on its value. Therefore an estimate of this 
 uncertainty can be simply obtained by computing the Bhabha cross section with $\Delta\alpha^{(5)}_{\rm hadr}(q^2)\pm\delta_{\rm hadr}$ and taking the
difference of the predictions as a measure of the associated theoretical error. One concludes that the spread between the minimum/maximum values and the central ones as returned by the two routines well agrees and does not exceed $2\times 10^{-4}$
at $\phi$ factories and $3\times 10^{-4}$ at $B$ factories \cite{Actis:2009gg}. This can be 
simply understood in terms of the dominance of $t$-channel photon exchange for large-angle Bhabha scattering at flavour factories and the 
excellent agreement between the two parameterizations for space-like virtualities. 

The total theoretical uncertainty of the most precise luminosity codes 
can be obtained by summing linearly the different 
sources of error discussed above, as shown in Table~\ref{tab1}. As can be seen, the total theoretical error 
amounts to about 0.1\% at $\phi$, $\tau$-charm and $B$ factories \cite{Actis:2009gg}. 

\begin{center}
\tabcaption{ \label{tab1} Summary of the different
sources of theoretical uncertainty for the most precise generators used
for luminosity measurements at flavour factories and the corresponding total
theoretical error for the calculation of the large-angle Bhabha cross section. From 
\cite{Actis:2009gg}.}
\footnotesize
\begin{tabular*}{80mm}{c@{\extracolsep{\fill}}ccc}
\toprule Source of error (\%) & $\phi$   & $\tau$-charm  & $B$ \\
\hline
$\,\,\,\,\,\,\, |\delta^{\rm err}_{\rm VP}|$~\cite{Jegerlehner:2006ju} & 0.00 &0.01 \hphantom{0} & 0.03 \\
$\,\,\,\,\,\,\, |\delta^{\rm err}_{\rm VP}|$~\cite{Hagiwara:2003da} &  0.02 &$\!\!\!\!\!\!\!$0.01 & \hphantom{0} $\!\!\!\!\!\!\!$0.02 \\
$|\delta^{\rm err}_{\rm SV}|$ \hphantom{00} & 0.02 & 0.02 \hphantom{0} & 0.02 \\
$|\delta^{\rm err}_{\rm HH}|$ \hphantom{0} & 0.00& 0.00 \hphantom{0} & 0.00 \\
$|\delta^{\rm err}_{\rm SV,H}|$ & 0.05 & 0.05 \hphantom{0} & 0.05\\
$|\delta^{\rm err}_{\rm pairs}|$ & 0.05 & 0.1 \hphantom{0} & 0.02\\
$|\delta^{\rm err}_{\rm total}|$ & 0.12$\div$0.14 & 0.18 \hphantom{0} & 0.11$\div$0.12\\
\bottomrule
\end{tabular*}
\end{center}

\section{Conclusions}

Stimulated by important experimental progress in the measurement of the low-energy hadronic cross section worldwide, the theoretical contribution to the uncertainty 
in the luminosity 
monitoring at flavour factories was significantly reduced in the last years down to the one per mille level. 

From one side, the development of new generators for the simulation of the Bhabha process (BabaYaga@NLO and MCGPJ) and the adoption of 
the LEP/SLC tool BHWIDE paved the way to theoretically consistent comparisons between the
predictions of these codes for all the observables measured by the experiments. The consistency stems from the fact that all these three generators include, albeit according to different details, the strictly necessary theoretical ingredients
given by NLO QED corrections matched with resummation, together with the contribution of vacuum polarization. 
Extensive work on tuned comparisons 
between the results of the luminosity tools greatly benefited of the activity of the WG 
on ``Radiative corrections and MC generators for low energies'' \cite{Actis:2009gg} and led to the conclusion that the large-angle Bhabha predictions of 
BabaYaga@NLO, BHWIDE and MCGPJ 
typically agree within $\sim$ 0.1\% for integrated cross sections and 
some per mille for distributions. These results, that take into account realistic selection criteria, demonstrate that the technical precision of these programs in their implementation of the radiative corrections is accurately under control, with a precision definitely 
sufficient in comparison with the present experimental errors. Actually, all these three codes are employed 
by the experimentalists to 
precisely monitor the luminosity of flavour factories, generally using more than one tool.

From the other side, various detailed cross-checks between the results of the generators and those of the presently available calculations of 
the NNLO QED corrections 
to the Bhabha cross section allowed to assess the theoretical 
accuracy of the MC tools on solid grounds. Indeed, the luminosity generators include, by construction, contributions of $O(\alpha^2)$ only approximately. 
Therefore, the comparison between the $O(\alpha^2)$ predictions of the generators and those of the exact 
perturbative calculations provides a reliable evaluation of the theoretical accuracy of the codes. The sum of the various sources of uncertainty (e.g. 
NNLO photonic contributions, lepton and hadron pair corrections, one-loop corrections to hard bremsstrahlung) is 
presently at the level of one per mille, again when considering selection cuts of actual experimental interest. This error estimate appears to be quite robust and is
presently dominated, as shown in Table~\ref{tab1}, by the uncertainties associated to the MC calculation of pair corrections (amounting 
to a few units in 10$^{-4}$) and of $e^+ e^- \to e^+ e^- \gamma$ with NLO accuracy (at the level of $\sim 5 \times 10^{-4}$). 
It is comparable to that achieved about a decade ago 
in the luminosity measurement at LEP/SLC colliders through small-angle Bhabha scattering \cite{Arbuzov:1996eq}.

In spite of this remarkable progress, further advances would be desirable. For example, the work on tuned comparisons was mainly made for 
the Bhabha process. It would be worthwhile to put forward such studies to cover the other luminosity reactions of interest, by comparing, for instance, the precision predictions of 
BabaYaga@NLO and MCGPJ for 
two photon production \cite{Balossini:2008xr,Fetodovich:2009} and the results of the whole set of the available programs
for the simulation of the processes of $\mu^+\mu^-$ and $\mu^+\mu^-\gamma$ production. For the latter processes some results can be found in 
\cite{Jadach:2005gx,Fetodovich:2009}. A better assessment of the theoretical accuracy could be achieved along the following directions. The preliminary analysis of the effect of lepton pair corrections 
performed in \cite{Actis:2009gg} should be extended to the study of hadron pairs and to a more thorough investigation of the pair correction 
dependence from the cuts of experimental interest. On this topic, work is already in a rather good shape but still in progress \cite{Carloni:2010}.
The results of the generators for the one-loop corrections to the 
radiative Bhabha signature should be compared with the predictions of the now available exact calculation 
\cite{Actis:2009uq}, again in the presence of realistic selection criteria. Such an analysis would be also important for the normalization process 
$e^+ e^- \to \mu^+\mu^-\gamma$. This study is ongoing but is still at a very preliminary stage \cite{Actis:2010}. 
Furthermore, following the same strategy used for Bhabha scattering, it would be also interesting to quantify 
to what extent the presently available parameterizations of $\Delta\alpha^{(5)}_{\rm hadr}(q^2)$, 
that manifest some discrepancies in their predictions for time-like momenta \cite{Teubner:2010ah,Actis:2009gg}, affect the uncertainty of the calculation of the 
cross section of $s$-channel annihilation into $\mu^+\mu^-$ and $\mu^+\mu^-\gamma$.

All these advances would eventually lead to improvements or refinements in the luminosity codes and possibly allow 
a control of the theoretical error in luminosity measurements at 
present flavour factories and at future super-$B$ colliders below the one per mille barrier.
\\

\acknowledgments{Guido Montagna thanks the organizers for the kind invitation and the very friendly atmosphere 
of a perfectly organized and interesting workshop.}

\end{multicols}

\vspace{-2mm}
\centerline{\rule{80mm}{0.1pt}}
\vspace{2mm}

\begin{multicols}{2}

\end{multicols}

\clearpage

\end{document}